\documentclass{amsbook}

\usepackage[T1]{fontenc}
\usepackage{ae,aecompl}

\usepackage{amssymb,amscd}
\usepackage{url}
\usepackage{amsthm,amsfonts,amsmath,amsxtra}
\usepackage{graphicx}
\numberwithin{equation}{section}
\numberwithin{table}{section}
\numberwithin{figure}{section}
\numberwithin{section}{chapter}

\newtheoremstyle{bold}
{.5\baselineskip}{.5\baselineskip}{\itshape}{}{\bfseries}{.}{.5em}{}
\newtheoremstyle{shy}
{.5\baselineskip}{.5\baselineskip}{}{}{\bfseries}{.}{.5em}{}

\makeatletter
\def\@captionfont{\small}
\def\mychapter{%
  \if@openright\cleardoublepage\else\clearpage\fi
 \thispagestyle{empty}\global\@topnum\z@
  \@afterindenttrue \secdef\@mychapter\@schapter}

\def\@mychapter[#1]#2#3{\refstepcounter{chapter}%
  \ifnum\c@secnumdepth<\z@ \let\@secnumber\@empty
  \else \let\@secnumber\thechapter \fi
  \typeout{\chaptername\space\@secnumber}%
  \def\@toclevel{0}%
  \ifx\chaptername\appendixname \@tocwriteb\tocappendix{chapter}{#2\\ \scshape #3}%
  \else \@tocwriteb\tocchapter{chapter}{#2\\ \scshape #3}\fi
  \chaptermark{#1}%
  \addtocontents{lof}{\protect\addvspace{10\p@}}%
  \addtocontents{lot}{\protect\addvspace{10\p@}}%
  \@mymakechapterhead{#2}{#3}\@afterheading}

\def\@mymakechapterhead#1#2{\global\topskip 7.5pc\relax
  \begingroup
  \fontsize{\@xivpt}{18}\bfseries\centering
    \ifnum\c@secnumdepth>\m@ne
      \leavevmode \hskip-\leftskip
      \rlap{\vbox to\z@{\vss
          \centerline{\normalsize\mdseries
              \uppercase\@xp{\chaptername\ \thechapter}}
          \vskip 3pc}}\hskip\leftskip\fi
     #1\par \vskip 1pc
     \Large\mdseries\scshape\centering
     #2\par \endgroup
  \skip@34\p@ \advance\skip@-\normalbaselineskip
  \vskip\skip@ }

\def\section{\@startsection{section}{1}%
  \z@{.9\linespacing\@plus\linespacing}{.5\linespacing}%
  {\large\bfseries\boldmath\centering}}

\def\subsection{\@startsection{subsection}{2}%
  \z@{.7\linespacing\@plus\linespacing}{.5\linespacing}%
  {\normalfont\scshape\centering}}

\def\theindex{\@restonecoltrue\if@twocolumn\@restonecolfalse\fi
  \columnseprule\z@ \columnsep 35\p@
  \@indextitlestyle
  \thispagestyle{empty}%
  \let\item\@idxitem
  \parindent\z@  \parskip\z@\@plus.3\p@\relax
  \raggedright
  \hyphenpenalty\@M
  \footnotesize}

\renewcommand{\@bibtitlestyle}{%
  \@xp\section\@xp*\@xp{\bibname}%
}
\renewcommand{\tocchapter}[3]{%
  \indentlabel{\@ifnotempty{#2}{\ignorespaces#1 #2.\quad}}#3}

\renewcommand{\tocsection}[3]{%
  \indentlabel{\@ifnotempty{#2}{\makebox[3.2em][l]{\ignorespaces#1 #2.}}}#3}

\makeatother

\renewcommand{\tocappendix}[3]{%
  \indentlabel{#1.\quad}#3}
\renewcommand{\tocappendix}[3]{%
  \indentlabel{\makebox[5.7em][l]{\ignorespaces#1.}}#3}

\renewcommand{\bibname}{References}

\renewcommand{\geq}{\geqslant}

\renewcommand{\leq}{\leqslant}

\theoremstyle{bold}
\newtheorem{theorem}{Theorem}[section]

\theoremstyle{shy}
\newtheorem{definition}[theorem]{Definition}

\newcommand{\cC}{\mathcal{C}}

\newcommand{\cG}{\mathcal{G}}

\newcommand{\cO}{\mathcal{O}}

\newcommand{\cT}{\ts\mathcal{T}}

\newcommand{\EE}{\mathbb{E}}

\newcommand{\VV}{\mathbb{V}}

\newcommand{\dd}{\ts\mathrm{d}\ts}

\newcommand{\ts}{\hspace{0.5pt}}

\newtheorem{result}[theorem]{Result}
\newcommand{\prob}{\mathrm{Prob}}
\newcommand{\cov}{\mathrm{Cov}}
\newcommand{\LamO}{\overset{\scriptstyle\circ}{\Lambda}{}}

\makeindex

\begin{document}

\mychapter{Counting, grafting and evolving binary trees}{Thomas~Wiehe$^*$}
\label{chap:TW}

\footnote{$^*$Institut f\"ur Genetik, Universit\"at zu K\"oln, Germany. Email: twiehe@uni-koeln.de}
Binary trees are fundamental objects in models of evolutionary biology
and population genetics. Here, we discuss some of their combinatorial
and structural properties as they depend on the tree class
considered. Furthermore, the process by which trees are generated
determines the probability distribution in tree space. \textit{Yule
  trees}, for instance, are generated by a pure birth process. When
considered as unordered, they have neither a closed-form enumeration
nor a simple probability distribution. But their ordered siblings have
both. They present the object of choice when studying tree structure
in the framework of evolving genealogies.

\section[Introduction]{Introduction}
Trees appear in different contexts and with different properties.  In
graph theory, they are defined as connected, acyclic graphs: any pair
of vertices (\textit{nodes}) is connected by exactly one concatenated
sequence of edges (\textit{branches}). Tagging one node, called
\textit{root} of the tree, implicitly establishes a directionality of
the graph.  In theoretical biology, trees are used to describe
genealogies of cells, genes, individuals or species. Depending on the
biological context, planarity of the tree, degree and labelling of
nodes, directionality and length of branches may or may not be of
interest. Cardinality and probability distribution depend
strongly on these properties.

The study of trees as mathematical objects reaches back at least to
the $1850$s, when Cayley \cite{TW-cay:1857} derived recursion formulas
for the enumeration of trees with a finite number of nodes, and also
recognised the link to isomer chemistry.  As an alternative to
recursions, bijections between trees and permutations can help to
solve certain counting problems \cite{TW-and:1881,TW-don:1975}.  More
generally, and yielding insight into asymptotic behaviour for large
trees, the tools of analytic combinatorics are particularly powerful.
Comprehensive treatments are found in the classical textbook by
Flajolet and Sedgwick \cite{TW-fls:09} and, focusing on random trees
only, in the textbook by Drmota \cite{TW-drm:09}.  With a view from
computer science, where they appear primarily as data structures,
trees are covered in the epitomic opus by Knuth
\cite[Vol.~1]{TW-knu:04}.

The link of `tree theory' with biology has been established by Yule's
seminal paper of 1925 \cite{TW-yul:1925}, when seeking to explain the
distribution of the number of species within genera. It initiated a
long tradition of research in phylogenetics and macro-evolution on
enumeration, topology and distribution of trees generated by random
processes
\cite{TW-cse:1967,TW-fel:1978,TW-slg:1989,TW-mas:1991,TW-kis:1993,TW-moh:1997,TW-sta:02,TW-blf:05,TW-las:13}.
The border between macro- and micro-evolution is fuzzy, but intensely
investigated in the context of gene tree embeddings in species trees
\cite{TW-ros:07,TW-drs:12,TW-las:13,TW-dir:17}.  Perhaps the most
genuine application of Yule's original model, and with most
ramifications, lies in population genetics as a model of individual
gene genealogies and their statistical properties.  Kingman's
\cite{TW-kin:1982} coalescent is its backward-in-time analogue and ---
in the guise of its evolved descendants --- features in several
chapters of this volume. The genetic operation of recombination
translates into subtree-prune and -regraft operations, opening a field
of active theoretical research on tree transformations
\cite{TW-son:06}, in part also covered in this volume.  Standard
references on the coalescent are the textbooks by Wakeley~\cite{TW-wak:09}
and Durett~\cite{TW-RD2008}.  Aldous \cite{TW-ald:01} offers a view on Yule's
paper from a modern perspective.

Given that trees are treated in different disciplines, and with
different degree of mathematical rigour, it is not surprising to find
oneself confronted with a non-unified, sometimes even inconsistent,
terminology and nomenclature, which alone can make it difficult to
identify the relevant theoretical features of some tree class for a
specific biological application.  Without claiming to authoritatively
clarify this problem, we start the section below with an (incomplete)
catalogue of tree classes and their enumerations
(Section~\ref{TW-section:counting}).  We will then devote special
attention to Yule trees and explore some of their structural
properties
(Sections~\ref{TW-section:ranking}~and~\ref{TW-section:inducing}). Since
they represent the scaffold of the widely used coalescent model in
population genetics, we will consider two such applications
(Sections~\ref{TW-section:recombining}~and~\ref{TW-section:evolving}).

\section[Counting trees]{Counting trees}\label{TW-section:counting}
\subsection[Preliminaries]{Preliminaries}
We consider rooted\index{tree!rooted}, binary\index{tree!binary},
finite trees: there is a unique node, the root, defining a
directionality for all branches. Each branch is delimited by a
\textit{parent} and a \textit{child} node. The root is
\textit{ancestor} of all other nodes. They are subdivided into
$n<\infty$ \textit{external} and $m=n-1$ \textit{internal} nodes,
including the root. All internal nodes have exactly two
children. External nodes have no \textit{descendants} and are also
called \textit{leaves}. The \textit{size}\index{tree!size} %\index{leaf}
of a tree is the number of its leaves. A
\textit{subtree}\index{subtree} is a tree that is rooted at some node
of the original tree.  Subtrees of size $2$ are also called
\textit{cherries}\index{cherry}, subtrees of size $3$
\textit{pitchforks}\index{pitchfork}. A
\textit{caterpillar}\index{caterpillar} is a tree for which at least
one of the subtrees at each of its internal nodes has size
$1$. Slightly more generally, a $c$\textit{-caterpillar} is a
(sub-)tree of size $c$ that is a caterpillar. Thus, a cherry is a
$2$-caterpillar, and a pitchfork is a $3$-caterpillar.  Since trees
here are binary, all internal nodes have a \textit{left} and a
\textit{right subtree}, which are rooted at the left and right
child. Trees are \textit{ordered}\index{tree!ordered}
(plane)\index{tree!plane}, if left and right can be distinguished,
otherwise they are \textit{un-ordered}\index{tree!unordered}
(non-plane).

\subsection[Classification]{Classification of binary trees}
Tree enumerations depend crucially on the presence and the kind of
node labels. Among the many possibilities, we restrict ourselves to
the following cases: presence or absence of alphanumeric labels at
external nodes, and presence or absence of totally ordered numeric
labels at internal nodes.  Trees without any node labels are called
\textit{shape trees} or
\textit{topologies} \cite{TW-cse:1967,TW-ros:06}.  We
call a tree \textit{ranked}\index{tree!ranked} or a \textit{history}
\cite{TW-har:1971,TW-smk:01,TW-drs:12}, if the internal nodes are
labelled with integers ${1,\dots , n-1}$ such that (i) the root has label
$1$, (ii) distinct nodes have distinct labels and (iii) every child
has a larger label than its parent.  We call a tree
\textit{labelled}\index{tree!labelled}, if the leaves carry labels.
Labelled trees can be thought of as phylogenies\index{phylogeny} with
species names as leaf labels. Without internal labels, they are also
called \textit{cladograms}, with internal labels they are
\textit{ranked phylogenies}\index{phylogeny!ranked} or
\textit{labelled histories}\index{history!labelled}
\cite{TW-smk:01}. Their cardinality follows, for instance, from a
coalsecent-like construction: randomly selecting two out of $k$
labelled lineages to coalesce, there are ${k \choose 2}$ possibilities
\cite{TW-las:13}. The product is
$\prod_{k=2}^{n} {k \choose 2} = n!(n-1)!/2^{n-1}$.
 
\begin{figure}[t]
\includegraphics*[angle=0,scale=.4,trim=0 0 0 30,clip]{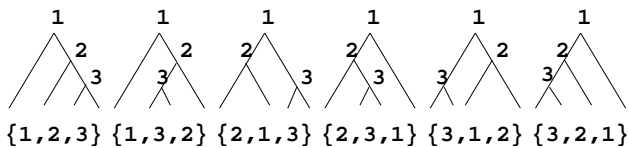}
\caption{\label{TW-fig:T4} The six ordered ranked trees of size $n=4$
  and the corresponding six permutations of $\{1,2,3\}$ obtained by
  reading out internal labels during \textit{in-order} tree
  traversal\index{tree!traversal} \cite{TW-knu:04}. Note, for example, the
  difference between $\{2,1,3\}$ and $\{3,1,2\}$.}
\end{figure}

When shape trees have a left/right orientation, they are called
\textit{Catalan}\index{tree!Catalan} trees, because they are
enumerated by the Catalan\index{Catalan number} numbers
$C_{m} = {2m \choose m}/(m+1)$, \cite[A000108]{TW-slo:1995}, where
$m=n-1$ is the number of internal nodes of such trees.  Finally,
\textit{ordered histories} are ordered ranked trees.  Since they map
bijectively to permutations\index{permutation} of $m=n-1$ integers, we
also call them \textit{permutation trees}\index{tree!permutation}.
They are enumerated by the factorials $m!$.  To see this, one can read
the labels of all ordered ranked trees of a given size in an
\textit{in-order} \cite{TW-knu:04} tree traversal, observing that all
subtrees, except cherries, have a distinguishable left-right order
(Figure~\ref{TW-fig:T4}).

We denote ordered trees of size $n$ by $\LamO_n^{..}$ and un-ordered
trees by $\Lambda_n^{..}$. The exponent is a placeholder to indicate
presence or absence of internal or external labels.  The tree classes
mentioned above are summarised in Table~\ref{TW-table:0}.
\begin{table}[ht]\caption{Classes of un-ordered ($\Lambda$) and
    ordered ($\LamO$) trees of size $n$. Presence ($+$) or absence
    ($-$) of internal or external labels is indicated by
    superscripts. Cardinalities are $|\Lambda_n|$ and
    $|\LamO_n|$.}\label{TW-table:0}
\begin{tabular}{l@{\hspace{2mm}}l@{\hspace{1mm}}c@{\hspace{2mm}}c@{\hspace{3mm}}c@{\hspace{2mm}}c@{\hspace{2mm}}c}
\hline\hline
name & alias & int. & ext. & symbol & cardinality & OEIS$^1$ ID\\
 & & lab. & lab. & & \\[5pt]
\multicolumn{7}{c}{unordered trees}\\ \cline{1-7} \\[-4pt]
shape trees & topologies$^2$ & $-$ & $-$ & $\Lambda_n^{--}$ & Eq.~(\ref{TW-eqlabel:e1}) & A001190\\[4pt]
ranked trees & histories$^3$ & $+$ & $-$ & $\Lambda_n^{+-}$ & Eq.~(\ref{TW-eqlabel:e2}) & A000111\\[4pt]
labelled trees & phylogenies$^4$ & $-$ & $+$ & $\Lambda_n^{-+}$ & $\frac{(2n-3)!}{2^{n-2}(n-2)!}$ & A001147\\[4pt]
  \begin{minipage}{59pt}labelled ranked trees$^5$\end{minipage} &
 \begin{minipage}{51pt}ranked phylogenies\end{minipage} & $+$ & $+$ & $\Lambda_n^{++}$ & $\frac{n!(n-1)!}{2^{n-1}}$ & A006472 \\[7pt]
  \multicolumn{7}{c}{ordered trees}\\ \cline{1-7} \\[-4pt]
  Catalan trees$^6$ & \begin{minipage}{51pt}ordered topologies\end{minipage} & $-$ & $-$ & $\LamO_n^{--}$ & $\frac{1}{n}{2(n-1) \choose (n-1)}$ & A000108\\[8pt]
  permutation trees & \begin{minipage}{49pt}ordered histories$^7$\end{minipage} & $+$ & $-$ & $\LamO_n^{+-}$ & $(n-1)!$ & A000142\\[8pt]
  \hline\hline
  \multicolumn{7}{l}{\footnotesize{$^1$ \url{www.oeis.org}, \cite{TW-slo:1995}}}\\[0pt]
  \multicolumn{7}{l}{\footnotesize{$^2$ \cite{TW-cse:1967,TW-ros:06}; called \textit{topological types} in \cite{TW-slg:1989}}}\\[0pt]
  \multicolumn{7}{l}{\footnotesize{$^3$ \cite{TW-har:1971,TW-smk:01,TW-drs:12}}}\\[0pt]
  \multicolumn{7}{l}{\footnotesize{$^4$ \cite{TW-ald:01,TW-slg:1989}; called \textit{rooted} phylogeny in \cite{TW-fel:1978} or \textit{tree form} in \cite{TW-cse:1967}}}\\[0pt]
  \multicolumn{7}{l}{\footnotesize{$^5$ cf. \cite{TW-las:13}, there in the context of Kingman's coalescent}}\\[0pt]
  \multicolumn{7}{l}{\footnotesize{$^6$ \cite[p. 5]{TW-drm:09}}}\\[0pt]
  \multicolumn{7}{l}{\footnotesize{$^7$ called \textit{shapes} in \cite{TW-har:1971}}}\\[-10pt]
\end{tabular}
\end{table}  
Note that these classes represent only a subset of the
possibilities. For instance, Felsenstein \cite{TW-fel:1978} discusses
phylogenies with non-numeric labels at internal nodes. This
constitutes a class that is different from $\Lambda_n^{++}$ and that 
has a different cardinality: it leads to Cayley's formula
\cite{TW-cay:1889}, enumerating non-binary trees
(cf. \cite[A000169]{TW-slo:1995} and \cite{TW-har:1971}).  Not all
tree classes have closed form enumerations. Often, ordered trees do,
while un-ordered trees do not \cite[p. 87]{TW-fls:09}.  In our list
(Table~\ref{TW-table:0}), the cardinalities of un-ordered shape and
ranked trees are given only implicitly via generating
functions\index{generating function}, but their ordered versions have
closed formulae.
 
The \textit{(ordinary) generating function}\index{generating function}
and the \textit{exponential generating function} of an integer sequence
$(a_n)_n$ are given by the formal power series
\[%\begin{array}{rclcrcl}
%f(x) &=& \sum_{n\geq 0} a_n x^n &\mbox{\ \ and\ \ }&
%F(x) &=& \sum_{n\geq 0} a_n \frac{x^n}{n!}\,,
% \end{array}
          f(x)=\sum_{n\geq 0} a_n x^n \qquad \text{ and } \qquad
          F(x)=\sum_{n\geq 0} a_n \frac{x^n}{n!}\, ,       
\]
respectively. If $f$ or $F$ are holomorphic functions defined in a
neighbourhood around $x=0$, the series can be interpreted as their
Taylor expansions and, for instance, their asymptotic properties can
be studied by analytic means.

In 1922, Wedderburn \cite{TW-wed:1922} showed that the cardinalities
of shape trees can be implicitly represented via a functional equation
of a generating function. De Bruijn and Klarner derived the somewhat
simpler representation
\begin{equation}
f(x) = x+ 1/2\left( f^2(x)+f(x^2)\right) \label{TW-eqlabel:e1}
\end{equation}
and showed \cite{TW-dbk:1982} that its solution $f$ generates the
cardinalities of shape trees of size $n$, via
$$
f(x) = \sum_n |\Lambda^{--}_n|\, {x^n}\,.
$$
For $1\leq n \leq 10$, the coefficients are $1, 1, 1, 2, 3, 6, 11, 23, 46, 98$.

For unordered ranked trees\index{tree!ranked}
(histories\index{history}), the cardinalities are identical with the
\textit{Euler numbers} and are given by the coefficients of the
exponential generating function
 \begin{equation}
F(x) = \sec(x) + \tan(x) = \sum_n |\Lambda^{+-}_{n+1}|\, \frac{x^n}{n!}\,,\label{TW-eqlabel:e2}
\end{equation}
for $1\leq n \leq 10$, they are $1, 1, 1, 2, 5, 16, 61, 272, 1385, 7936$.

A natural way to construct unordered ranked trees of any finite size
is by recursion: given a ranked tree of size $m=n-1$, construct a tree
of size $n$ by randomly choosing one of the $m$ leaves to give rise to
two children and label the chosen leaf with the integer $n$.
Following other authors \cite{TW-smk:01,TW-cf:10}, we call trees
generated in this way \textit{Yule trees}\index{tree!Yule} and the
underlying model (process) the \textit{Yule model} (\textit{Yule
  process})\index{process!Yule}\index{Yule process}. In the equivalent backward process,
one starts from $n$ leaves and their $n$ parental branches. One
randomly, and iteratively, selects two branches to coalesce into a
single one until all are coalesced. When, in addition, a time axis for
the coalescent times is introduced, and when these times are
exponentially distributed with a parameter proportional to
$k\choose 2$, where $k$ is the current number of branches, Yule trees
are called \textit{coalescent trees},\index{tree!coalescent}\index{coalescent!tree}
generated by the (Kingman-) coalescent process~\cite{TW-kin:1982}.\index{process!coalescent}\index{coalescent!Kingman} They are the basis of a plethora of genealogical models in population genetics.

\begin{figure}[t]
\begin{center}
\begin{tabular}{c|ccccc}\hline\hline ~\\[-9pt]
ranked trees & $|\cC_2|$ & $|\cC_3|$ & $|\Lambda|$ & factor & $|\LamO|$ \\ \hline
\includegraphics*[angle=0,scale=.319,trim=0 0 0 0]{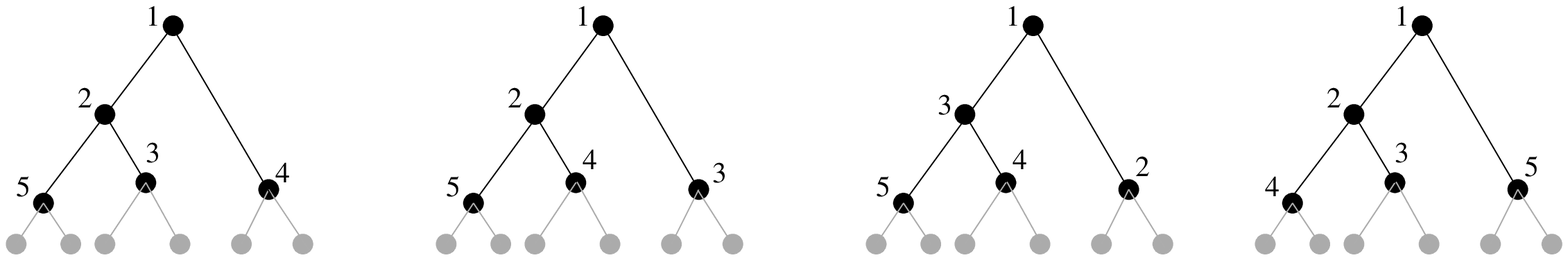} & 3 & 0
& $4$ & $2^{5-3}$& $16$ \\ \hline
\includegraphics*[angle=0,scale=.319,trim=0 0 0 0]{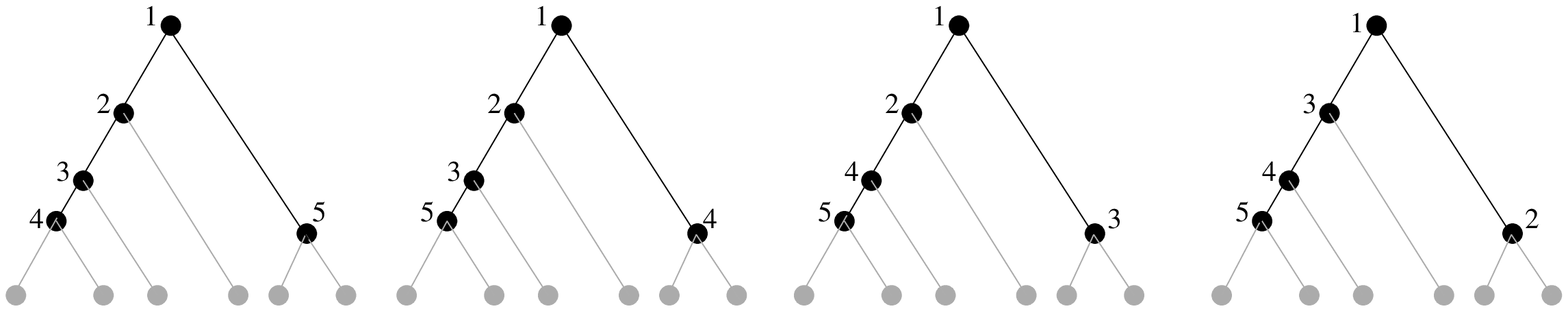} & 2 & 1
& $4$ & $2^{5-2}$ & $32$ \\ \hline
\includegraphics*[angle=0,scale=.319,trim=0 0 0 0]{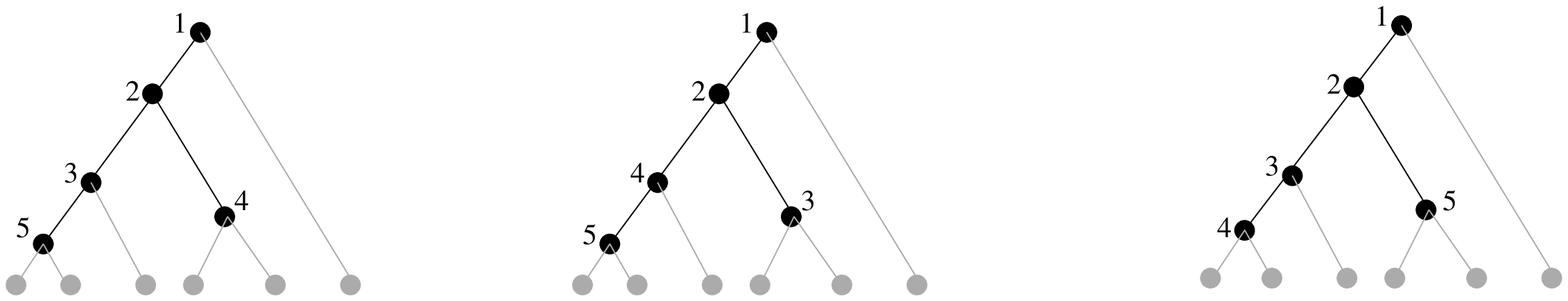} & 2 & 1
& $3$ & $2^{5-2}$ & $24$\\ \hline 
\includegraphics*[angle=0,scale=.319,trim=0 0 0 0]{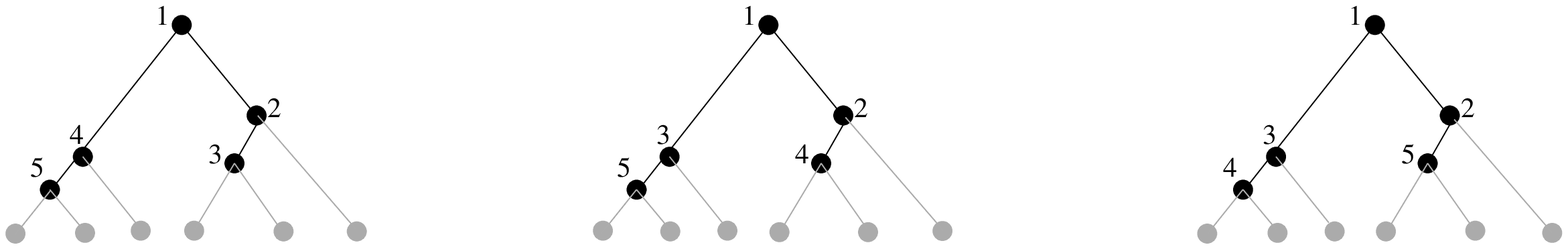} & 2 & 2
& $3$ & $2^{5-2}$ & $24$\\ \hline
\includegraphics*[angle=0,scale=.319,trim=0 0 0 0]{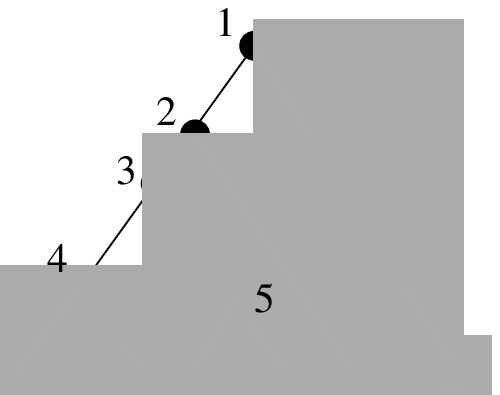} & 2 & 0
& $1$ & $2^{5-2}$ & $8$\\ \hline
\includegraphics*[angle=0,scale=.319,trim=0 0 0 0]{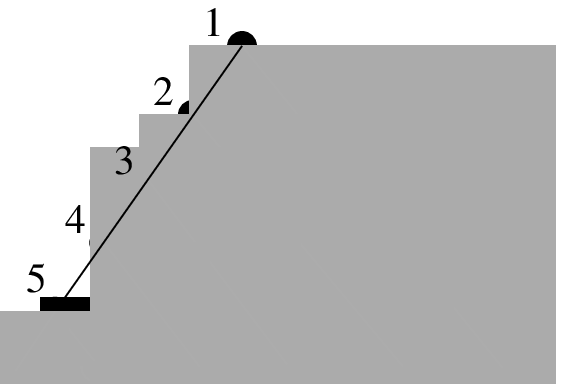} & 1 & 1
& $1$ & $2^{5-1}$ & $16$\\ \hline\hline
\end{tabular}
\end{center}
\caption{\label{TW-figlabel:f3} The sixteen possible un-ordered ranked
  trees\index{tree!ranked} of size $n=6$, classified by shape. Within
  each class, all admissible orderings of the internal nodes are
  displayed. Number of cherries ($|\cC_2|$) and pitchforks ($|\cC_3|$)
  are indicated. The number of all ordered ranked trees, classified by
  shape, is obtained by multiplying with the factor
  $2^{m-|\cC_2|}$. The total number is $5!=120$. Branch lengths are
  without meaning; position of an internal node in a tree is given by
  the node label, not by the actual drawing of its position. External
  nodes and branches are shown in grey. Removing them leads to the
  \textit{reduced trees}\index{tree!reduced} of size $5$. They can be
  uniquely identified with the original trees of size $6$.}
\end{figure}

\section{Properties of ranked trees}\label{TW-section:ranking}
Note that the Yule process does not generate uniformly distributed
trees in $\Lambda_n^{+-}$. For instance, in Fig~\ref{TW-fig:T4} the
$4$-caterpillar is generated with probability $2/3$ and the balanced
tree, corresponding to the permutations $\{2,1,3\}$ and $\{3,1,2\}$,
with probability $1/3$. Only when considered as trees in
$\LamO_n^{+-}$, they become uniformly distributed under the Yule
process, each with probability $1/(n-1)!$. Other tree generating
processes may lead to still other probability distributions
\cite{TW-mas:1991}.

Since ordered and un-ordered trees are identical up to left/right
order of subtrees that are not cherries, there are exactly
$ 2^{n-1-o} $ different ordered trees for each unordered one with $o$
cherries\index{cherry}.  Thus, given a ranked tree, one also knows the
probability with which it is generated, by simply counting its
cherries (cf. \cite{TW-taj:1983}). With $\cO$ denoting the random
variable for the number of cherries, we have
\begin{equation}\label{TW-eqlabel:e4}
  \prob (\mbox{given ranked tree of size $n$ with $\cO=o$ cherries})
  = \frac{2^{n-1-o}}{(n-1)!}\,.
\end{equation}
To explore the unconditional distribution of Yule trees, we remark
that all external nodes and branches (shown in grey in
Figure~\ref{TW-figlabel:f3}) may be stripped from a ranked tree of
size $n$ without loss of information. Such stripping leads to a
\textit{reduced}\index{tree!reduced} tree with $m=n-1$ nodes with
ordered labels, all of out-degree $0$, $1$ or $2$ \cite{TW-diw:13}.
Nodes of out-degree $0$ represent cherries in the original tree.
Sometimes, reduced trees are called \textit{pruned} trees
\cite{TW-fls:09}, a term which we avoid, to not confuse it with `tree
pruning' discussed later.  Reduced trees with $m$ nodes can be
constructed recursively, starting from a reduced tree with one node,
according to the following production rule
$$
(o,m) \longrightarrow (o,m+1)^o (o+1,m+1)^{m-2o+1}\,,
$$
where $o$ is the number of cherries and $m$ the total number of nodes
in the current tree. The exponent counts how many new trees with $o$
(or $o+1$) cherries and $m+1$ nodes are produced. Note that in each
step $m$ is increased by one and the number of cherries may either
remain unchanged or also increase by one.  The former happens when the
new branch and node are appended at a node of out-degree $0$, the
latter, when appended at a node of out-degree $1$. At nodes of
out-degree $2$ (true internal nodes) nothing can be appended.  For
instance, starting with $(1,1)$, the production rule generates the
sequence
$$
(1,2)^1 (2,2)^0,\  (1,3)^1 (2,3)^1,\ (1,4)^1 (2,4)^2
\mbox{ and } (2,4)^2 (3,4)^0, \dots \,.
$$
Consider now the bivariate exponential generating function
\begin{equation}\label{TW-equation:bivegf}
F(x,z) = \sum_{{\stackrel{\mbox{\tiny reduced trees with $o$}}{\mbox{\tiny cherries and $m$ nodes}}}} x^o\frac{z^m}{m!}\,.
\end{equation}
The production rule can then be translated into algebraic terms as
\begin{eqnarray*} 
  F(x,z)&=&xz + \sum \frac{o x^o z^{m+1}}{(m+1) \, !} + \sum
            \frac{(m-2o+1)(x^{o+1}z^{m+1})}{(m+1) \, !} \\
        &=& xz + (1-2x)\sum \frac{o x^o z^{m+1}}{(m+1) \, !} + xz \sum
            \frac{x^{o}z^{m}}{m \, !} \,,
\end{eqnarray*}
where the summations are over all reduced trees with $o$ cherries and
$m$ nodes and the first summand represents a tree of size $m=1$.
Differentiating both sides with respect to the variable $z$, one
obtains a partial differential equation for $F$
\begin{equation*}
  x(1-2x) \frac{\partial F}{\partial x}(x,z) + (xz-1)
  \frac{\partial F}{\partial z}(x,z) = -x F(x,z) - x\,,
\end{equation*}
which admits a solution in closed form \cite{TW-diw:13} as
\begin{equation}\label{TW-eqlabel:e5}
F(x,z)= \frac{2\big( x \, \exp(z \sqrt{-2x+1}) - x \big)}{(\sqrt{-2x+1}-1)\exp(z \sqrt{-2x+1}) + \sqrt{-2x+1} + 1}.
\end{equation}

One direct application of $F$ is to determine the probability that two
randomly generated Yule trees are identical (\cite[Thm.~1]{TW-diw:13},
with $F$ replaced by $Y$).  Furthermore, $F$ can be used to find a
\textit{partition}\index{partition} of the Euler\index{Euler number}
numbers $e_m$ in such a way that $e_{m,o}$ represents the number of
(unreduced) ranked trees of size $n=m+1$ with $o$ cherries. As shown
in \cite{TW-diw:13},
$$
e^{}_{m,o} = m! \cdot [x^o z^m] F\,,
$$
where the brackets $[\cdot]$ denote coefficient extraction.  The
partitions of $e_m$ for $m=1,\dots ,10$ and $o=1,\dots ,5$ are shown in
Table~\ref{TW-table:Eul}.
\begin{table}\caption{Partitions $e^{}_{m,o}$ of Euler numbers
    \cite[A000111]{TW-slo:1995}. $\cO$: number of cherries. Column
    sums $\sum_o e^{}_{m,o} = e^{}_m$. For instance, for $m=5$ (i.e., $n=6$)
    there are one ranked tree with one cherry (the caterpillar), $11$
    trees with two cherries and $4$ trees with three
    cherries.}\label{TW-table:Eul}
\begin{tabular}{rrrrrrrrrrr}
\hline\hline 
$\cO$  & \multicolumn{10}{c}{$m$}\\ \cline{2-11}
& $1$ & $2$ & $3$ & $4$ & $5$ & $6$ & $7$ & $8$ & $9$ & $10$ \\ \hline
$1$ &  1   & 1    &  1   & 1    &  1   &  1   & 1    & 1    & 1    & 1 \\ 
$2$ &  0   & 0    &  1   & 4    &  11  &  26  & 57   & 120   & 247  & 502 \\ 
$3$ &  0   & 0    &  0   & 0    &  4   &  34  & 180  & 768  & 2904 & 10194 \\ 
$4$ &  0   & 0    &  0   & 0    &  0   &  0   & 34   & 496  & 4288 & 28768 \\ 
$5$ &  0   & 0    &  0   & 0    &  0   &  0   &  0   &  0   & 496  & 11056 \\ 
$\sum$ &  1   & 1    &  2   & 5    &  16   &  61   &  272   &  1385   & 7936  & 50521 
\\\hline \hline
\end{tabular}
\end{table}
Other applications involve simple transformations of $F$. 
For instance, with
$$
{\tilde F}(x,z) = z F\big(\tfrac{x}{2},2z\big)
$$
one obtains the weighted (ordinary) generating function
\begin{equation}\label{TW-eqlabel:e6}
{\tilde F}(x,z)= \frac{zx \exp{\left(2z \, \sqrt{-x + 1}\right)} - zx}{{\left( \sqrt{-x  + 1} -1\right)} \exp{\left(2z \, \sqrt{-x  + 1}\right)} + 1 + \sqrt{-x  + 1}}\,,
\end{equation}
for the coefficients of 
$x^o z^n$, such that
\begin{equation} \nonumber {\tilde F}(x,z)=\sum_{\mbox{{\tiny ranked
        trees of size $n$}}} \frac{2^{n-1-o}}{(n-1)!}x^{o}z^{n}\,,
\end{equation}
leading to the following \cite{TW-diw:13} consequence.
\begin{result}
  The probability that a Yule tree of size $n$ has $o$ cherries is
  given by the coefficient of $x^{o}z^n$ in the Taylor expansion of
  ${\tilde F}$ around $z=0$, i.e.,
$$
P^{}_n(\cO=o)=[x^{o}z^n]{\tilde F}(x,z)\,.
$$
\end{result}
By differentiating $\tilde F$, one can easily derive the moments of $\cO$.
For instance, the mean number of cherries in ranked trees of size $n$ is
$$
\EE (\cO)=[z^n]\left.\frac{\partial{\tilde F}}{\partial
    x}(x,z)\right|^{}_{x=1}=[z^n]\frac{z^4-3z^3+3z^2}{3(z-1)^2}\,.
$$
If $n>2$, this simplifies to
$$
\EE (\cO)=\frac{n}{3}\,.
$$

The second moment is
\begin{eqnarray*} 
  \EE (\cO^2)&=&[z^n]\left.\frac{\partial(x\frac{\partial{\tilde F(x,z)}}{\partial x})}{\partial x}\right|^{}_{x=1}=[z^n]\left.
   \frac{\partial^2{\tilde F(x,z)}}{\partial x^2}\right|^{}_{x=1}+ [z^n]\left.
     \frac{\partial{\tilde F(x,z)}}{\partial x}\right|^{}_{x=1} \\
&=& [z^n]\left(\frac{2}{(z-1)^3}\left(\frac{z^7}{45}-\frac{2z^6}{15}+\frac{z^5}{3}-\frac{z^4}{3}\right)\right) + \EE  (\cO)\,.
\end{eqnarray*}
If $n>6$, and using $\VV (\cO)=\EE (\cO^2)-\EE^2(\cO)$, one obtains
$$
\VV (\cO) =\frac{2n}{45}\,.
$$

\begin{center}
    \begin{figure}[t]
      \includegraphics[scale=0.4,angle=0,trim=10 70 10 70,clip]{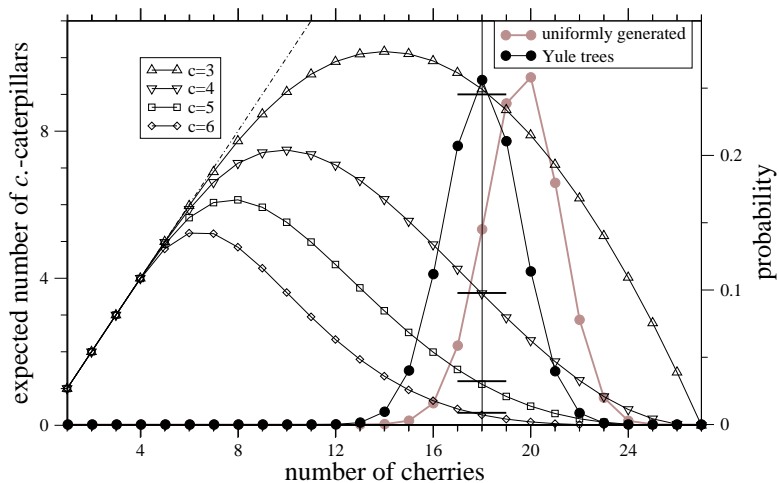}
\caption{\label{TW-figlabel:f7} Ranked trees\index{tree!ranked} of
  size $n=54$.  Conditional expectation of the number of
  $c$-caterpillars (left $y$-axis, $c=3,4,5,6$), given the number of
  cherries (curves with triangles, diamonds and squares). Vertical
  black line at $x=18$: expected number of cherries in unconstrained
  trees; horizontal black bars: unconditional expected number of
  $c$-caterpillars. Curves with filled circles: fraction of trees
  (right $y$-axis) with given number of cherries generated under the
  Yule process (black) and in uniformly generated trees (grey).
  Equivalently, this is the distribution of cherries ($\cO$) in ranked
  trees. $\VV (\cO)/\EE (\cO)\approx 0.13$. Dotted line: diagonal
  $x=y$.}
\end{figure}
\end{center}

The distribution of $\cO$ \cite{TW-mks:00}, and mean and variance of
$c$-caterpillars \cite{TW-ros:06}, have been derived before, however
with different methods not employing generating functions.  The latter
represent a powerful tool to handle the recursive production rules of
binary trees, and readily offer a somewhat deeper look into tree
structure. Focusing on general $c$-caterpillars\index{caterpillar},
let
$$
F(x^{}_2,x^{}_3,x^{}_4,\dots ,x^{}_k,z) = \sum_{\mbox{\tiny trees of size $n>1$}} 
x_2^o  x_3^{c^{}_3} x_4^{c^{}_4}\dots x_k^{c^{}_k} \frac{z^{n-1}}{(n-1)!} 
$$
    be a multi-variate exponential generating function\index{generating function},
    where $c_i$ is the number of caterpillars of size $i>2$,
    and $o$ the number of cherries.  This function satisfies the partial
    differential equation
\begin{eqnarray*}
  \frac{\partial{F}}{\partial z} &=& x^{}_2 + x^{}_2 {F} + x^{}_2 z \frac{\partial{F}}{\partial z} + (x^{}_2 x^{}_3 - 2 x_2^2)
                                     \frac{\partial{F}}{\partial x^{}_2}   \nonumber \\
                                 & & + \sum_{i=3}^{k-1} \Big( x^{}_i x^{}_{i+1} - x_i^2 + x^{}_2(1-x^{}_i)\Big( 1 + \sum_{j=1}^{i-3}
                                     \frac{1}{x^{}_{i-1}x^{}_{i-2}\dots x^{}_{i-j}}\Big) \Big)
                                     \frac{\partial{F}}{\partial x^{}_i}   \nonumber \\
                                 & &  + \Big(x^{}_k -x_k^2 + x^{}_2(1-x^{}_k)
                                     \Big( 1 + \sum_{j=1}^{k-3}\frac{1}{x^{}_{k-1}x^{}_{k-2}\dots x^{}_{k-j}}\Big) \Big)
                                     \frac{\partial {F}}{\partial x^{}_k}\,, \nonumber
\end{eqnarray*}
which leads to a recursively determined family of polynomials
$(F_m)_{m\geq 1}$ with
$$
F^{}_m= \sum_{\mathrm{trees\,} t \mathrm{\,of\,size\,} n=m+1} \frac{x_2^{o(t)}{x^{}_3}^{c_3(t)}{x^{}_4}^{c_4(t)}\dots x_k^{c_k(t)}z^{n-1}}{(n-1) !}\,.
$$
Defining the operator
$$
\cG(F) = \frac{\partial{F}}{\partial z} - x^{}_2\,,
$$
the recursion for $(F_m)_{m \geq 1}$ is given by
\begin{equation} \label{TW-eqlabel:e8} %\nonumber
  \begin{split}
    F^{}_1 &= x^{}_2 z\,, \\ %\label{TW-eqlabel:e8}
    F^{}_{m+1} &= \int \cG({F^{}_m})\,  \dd z\,. %\\ \nonumber
    \end{split}
\end{equation}
As an example, fix $k=5$. Then, for $m=1, 2, 3, 4, 5$, one has
\[ \begin{aligned} %\begin{eqnarray*} 
F^{}_1 = &\ x^{}_2 z \,, \\
F^{}_2 =&\ \frac{1}{2} x^{}_3 x^{}_2 z^2 \,,\\
F^{}_3 =&\ \frac{1}{6} x^{}_3 x^{}_4 x^{}_2 z^3 + \frac{1}{6} x_2^2 z^3  \,,\\
F^{}_4 =&\ \frac{1}{24} x^{}_3 x^{}_4 x^{}_5 x^{}_2 z^4 + \frac{1}{24} x_2^2 z^4 + \frac{1}{8} x^{}_3 x_2^2 z^4 \,,\\
F^{}_5 =&\ \frac{1}{120} x^{}_3 x^{}_4 x^{}_5 x^{}_2 z^5 + \frac{1}{120} x_2^2 z^5 + \frac{1}{40} x^{}_3 x_2^2 z^5 + \\
 & \frac{1}{40} {x_3}^2 x_2^2 z^5 + \frac{1}{30} x^{}_3 x^{}_4 x_2^2 z^5 + \frac{1}{30} x_2^3 z^5.\\
 % \end{eqnarray*}
                      \end{aligned}
 \]
Recursion (\ref{TW-eqlabel:e8}) yields both the joint distribution of
cherries and caterpillars of different sizes and the conditional
distribution of caterpillars, conditioned on the number of cherries.
Summarising, one can state the following result
(cf. \cite{TW-diw:13}).
\begin{result}\label{TW-theorem:pollo}
Given an (unordered) ranked tree $T$ of size $n=m+1$. Then,
\begin{itemize}
\item[i)] the probability that $T$ contains $c$-caterpillars of size
  $k$ is
 \[ 
  P^{}_m(\cC^{}_k= c) = [x_k^{c}]F^{}_m \Big( \frac{1}{2},1,1,\dots ,x^{}_k,2 \Big);
\]
\item[ii)] the joint probability that $T$ contains $o$ cherries and
  $c$ caterpillars of size $k$ is
\[
P^{}_m(\cO= o,\, \cC^{}_k= c) = [x_2^{o}x_k^{c}]F^{}_m
\left( \frac{x^{}_2}{2},1,1,\dots ,x^{}_k,2 \right);
\]
\item[iii)] the conditional probability that $T$ contains $c$
  caterpillars of size $k$, given it has $o$ cherries is
\[
P^{}_m (\cC^{}_k= c \,|\, \cO= o) = \frac{P^{}_m(\cO= o,\cC^{}_k= c)}{P^{}_m(\cO= o)}
= \frac{[x_2^{o}x_k^{c}]F^{}_m \left( \frac{x^{}_2}{2},1,1,\dots ,x^{}_k,2 \right)}{ [x_2^{o}]F^{}_m \left( \frac{x^{}_2}{2},1,1,\dots ,1,2 \right)};
\]
\item[iv)] the probability that $T$ contains $c'$ caterpillars of size
  $i$, with $3\leq i < k$, and $c$ caterpillars of size $k$ is
\[
P^{}_m(\cC^{}_i= c',\, \cC^{}_k= c) = [x_i^{c'}x_k^{c}]F^{}_m \Big( \frac{1}{2},1,\dots
  1,x^{}_i,1,\dots ,x^{}_k,2 \Big);
\]
\item[v)] the conditional probability that $T$ contains $c$
  caterpillars of size $k$, given it has $c'$ caterpillars of size
  $i$, with $3\leq i < k$, is
\begin{eqnarray} \nonumber
P^{}_m(\cC^{}_k= c \,|\, \cC^{}_i= c') &=& \frac{P^{}_m(\cC^{}_i= c',\cC^{}_k= c)}{P^{}_m(\cC^{}_i= c')} \\\nonumber
 &=& \frac{[x_i^{c'}x_k^{c}]F^{}_m \left( \frac{1}{2},1,\dots ,1,x^{}_i,1,\dots ,x^{}_k,
  2 \right)}{ [x_i^{c'}]F^{}_m \left( \frac{1}{2},1,\dots ,1,x^{}_i,1,\dots ,1,2 \right)}. \nonumber
\end{eqnarray}
\end{itemize}
\end{result}
The distribution of $\cO$, both under the Yule process and when trees
are generated uniformly, as well as the conditional expectations for
some $c$-caterpillars, are shown in Figure~\ref{TW-figlabel:f7}
for the example of size $n=54$.

\section[Induced subtrees]{Induced subtrees}\label{TW-section:inducing}

Induced subtrees\index{subtree!induced} occur as embedded genealogies
of a subset of the leaves of a tree \cite{TW-stw:1984}.  Let $T_n$ be
a ranked, labelled tree of size $n$ with leaf labels
$L=\{l_1,l_2,\dots ,l_n\}$.  Choose $n'\leq n$, and select labels
$L'=\{l'_1,l'_2,\dots ,l'_{n'}\}$, such that for each $1\leq i\leq n'$
there is exactly one $j$ with $l'_i=l_j$. Then, the \textit{induced
  subtree} $T'$ is the tree that is obtained from $T$ by maintaining
only the branches connecting a leaf $l'_i$ with the most recent
common ancestor of all leaves $L'$. We write $T'\lhd T$ for
short. Note that the root of $T'$ is not necessarily identical with
the root of $T$ and that the topologies of different induced subtrees
of the same supertree $T$ may be different.  There are $n \choose n'$
possible subsets of size $n'$.  When conditioned on a fixed tree $T$,
number and distribution of induced subtrees are obviously different
from independently generated trees. There is no general enumeration
formula for induced subtrees since the number depends on the topology
of $T$.  For instance, take a caterpillar of size $n$. Then all
induced subtrees are caterpillars.  Only when averaging over all Yule
super-trees of size $n$, induced subtrees and independently generated
trees are identical in number and distribution.  We introduce now the
notion of \textit{node balance}\index{balance!node}.

\begin{definition}
  For an internal node $\nu_i$ of a binary rooted tree $T$ let
  $T_i(L)$ and $T_i(R)$ be the left and right subtrees at node
  $\nu_i$. We call the minimum
\[
\omega^{}_i = \min\{|T^{}_i(L)|,|T^{}_i(R)|\}
\]
\textit{node balance} at node $\nu_i$.
In particular, $\omega_1$ is the \textit{root balance}\index{balance!root}.
\end{definition}

It is a standard exercise to calculate the probability that $T$ and
$T'$ have the same root ($\nu_1$). Given $T$ and fixing $\omega_1$,
one has
\begin{equation*}
\prob (\nu_1' = \nu^{}_1\,|\,T, \omega^{}_1) = \sum_{i=1}^{n'-1} \frac{ {\omega^{}_1 \choose i}{n-\omega^{}_1 \choose n'-i}}{{n \choose n'}}=1-\frac{ {\omega^{}_1 \choose n'}+{n-\omega^{}_1 \choose n'}}{{n \choose n'}}\,.
\end{equation*}
When $n$ is large, one may replace the hypergeometric terms by
binomials and get
\begin{equation}
  \prob(\nu_1' = \nu^{}_1\,|\,T, \omega^{}_1) \approx \sum_{i=1}^{n'-1}
  {n' \choose i} p^i (1-p)^{n'-i}
=1-(1-p)^{n'} - p^{n'}\,,\label{TW-eq:six}
\end{equation}
where $p=\omega_1/n$, $0<p\leq 1/2$.  For trees generated by the Yule
process, node balance is (nearly) uniformly distributed on
$1,\dots ,\lfloor n/2\rfloor$, hence $p$ is uniform on $]0,1/2[$.
Integrating Eq.~(\ref{TW-eq:six}) with respect to $p$ and multiplying
with uniform weights, one obtains the well known result
(cf. \cite{TW-stw:1984})
\begin{equation*}
  \prob(\nu_1' = \nu^{}_1) \approx 2 \int_0^{1/2} \left(1-(1-p)^{n'} - p^{n'}\right)
  \dd p = \frac{n'-1}{n'+1}\,.
\end{equation*}

We now consider node balance in induced subtrees.  Let the random
variable $\Omega_1$ be root balance in a Yule tree\index{tree!Yule} of
size $n$.  One has
\[
\mbox{Prob}(\Omega^{}_1=\omega^{}_1) = \frac{2- \delta^{}_{\omega^{}_1,n/2}}{n-1}\,.
\] 
Fixing $T$ and selecting an arbitrary induced subtree $T' \lhd T$,
consider the random variable $\Omega_1'\mid\Omega_1$. To calculate the
conditional distribution, one may use the auxiliary terms
\begin{eqnarray*}
  p(\omega'_1 \mid \omega^{}_1) & \approx & 
 \prob(\upsilon^{}_1=\upsilon'_1) \cdot \left(\frac{{{\omega^{}_1}\choose{\omega'_1}}{{n-\omega^{}_1}\choose{n'-\omega'_1}}+{{n-\omega^{}_1}\choose{\omega'_1}}{{\omega^{}_1 }\choose{n'-\omega'_1}}}{{{n}\choose{n'}} - {{\omega^{}_1}\choose{n'}} - {{n-\omega^{}_1}\choose{n'}}}\right) \left( \frac{1}{1+\delta^{}_{\omega'_1 , n'/2}}  \right) \\\nonumber
   && \ +\ \prob(\upsilon^{}_1 \neq \upsilon'_1) \cdot \left(\frac{2-\delta^{}_{\omega'_1,n'/2}}{n'-1}\right)\,, \nonumber 
\end{eqnarray*}
assuming that the induced subtree $T'$ is a random tree of size $n'$ 
when roots of $T$ and $T'$ are different. 
Normalising, one obtains 
\begin{equation}\label{TW-equation:appol}
  \prob(\omega_1' \mid \omega^{}_1) = \left( \sum_{\omega_1' = 1}^{\lfloor n'/2 \rfloor}
    p (\omega'_1 \mid \omega^{}_1) \right)^{-1} \cdot p (\omega'_1 \mid \omega^{}_1).
\end{equation}

Different roots, and the ensuing `approximation', are likely to occur
when $\omega_1$ is small.  Analytical, however lengthy, expressions of
the conditional expectation $E(\Omega'_1\mid\Omega_1)$ are then easily
derived with software for symbolic algebra.
 
\begin{center}
\begin{figure}[t]
 \includegraphics[scale=0.41,trim=10 40 0 80,clip]{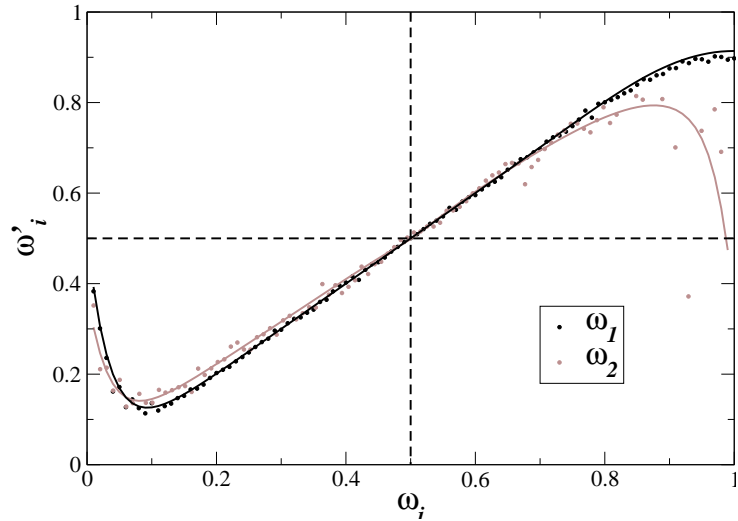}
 \caption{\label{TW-figure:tablew} Standardised (i.e., scaled to
   $[0,1]$) values of $\EE (\omega'_1 \mid \omega_1)$ (black) and
   $\EE (\omega_2' \mid \omega_2)$ (grey) for $n=200$ and
   $n'=50$. Theoretical results (solid lines) according to
   Eq.~(\ref{TW-equation:appol}) and simulation results (dots),
   obtained with {\tt ms} \cite{TW-hud:02}.}
\end{figure}
\end{center}

This computation can be extended to the balance $\Omega_2$ of the root
of the largest root\index{balance!root} subtree, to obtain the
conditional expectation of $\Omega'_2\mid(\Omega_1,\Omega_2)$ and of
$\Omega'_2\mid\Omega_2$ (Disanto and Wiehe, unpublished results).  In
Fig.~\ref{TW-figure:tablew}, we show $E(\Omega_1' \mid \omega_1)$ and
$E(\Omega_2' \mid \omega_2)$ as functions of $\omega_1$ and
$\omega_2$ and compare them to simulated values. Shown are averages
across arbitrary trees of fixed size $n$ and arbitrary induced
subtrees of fixed size $n'$.  Note that induced subtrees, when
conditioned on a fixed super-tree, reflect node balance of the
supertree only when the latter is not extremal.  In principle, these
calculations could be continued to further internal nodes.  However, a
full probabilistic treatment and the involved expressions become very
clumsy.

\subsection*{Application: Neutrality test using node balance}

Tree balance statistics \cite{TW-col:1982,TW-kis:1993,TW-blf:05} have
traditionally been used to investigate evolutionary hypotheses in the
context of phylogenetic species trees. However, they can also be
defined and examined for gene genealogies modelled by the coalescent
process and be integrated into powerful tests of the neutral evolution
hypothesis \cite{TW-li:11,TW-lw:13,TW-flwar:17}.  Published versions
of such tests, however, are typically a mixture of tree shape and
branch length statistics.  Relying, in contrast, only on node
balance\index{balance!node}, one may define the statistic
(cf. \cite{TW-lw:13})
\begin{equation}
{\cT}^{}_3 = 2 \sum_{i=1}^3 \big( 2 \tfrac{\Omega^{}_i}{n^{}_i} -\tfrac{1}{2}\big)\,,
\end{equation}
where $n_1=n$, $n_2=n-\Omega_1$ and $n_3=n-\Omega_1-\Omega_2$. Since
$2 \Omega_i/n_i$ is approximately uniform on the interval $[2/n_i,1]$,
${\cT}_3$ is close to standard normal \cite{TW-lw:13}. Small values of
${\cT}_3$ are obtained for highly unbalanced trees, i.e., when
$\omega_i$ are small, produced for instance by
caterpillars\index{caterpillar}, and large values for highly balanced
trees.  In the context of population genetics, a locally unbalanced
genealogy of a sample of $n$ genes can be produced by the rapid
fixation of a favourable allele. Hence, an estimate of $\cT_3$, based
on observed genetic variability, provides a statistic with which the
hypothesis of neutral evolution can be tested. The results on induced
subtrees can be integrated into a nested test-strategy where samples
and sub-samples are tested jointly. More details are described in
\cite{TW-rau:18}.

\section[Recombination]{Transformations I: Pruning, grafting and
  recombination}\label{TW-section:recombining}

Let $T\in \Lambda_n^{+-}$ be a ranked tree.  The \textit{layer} $l_j$
($1\leq j \leq n$) of $T$ is the `interval' in which $T$ has $j$
branches. Layer $l_1$ can be imagined as the infinitely long layer
above the root, which makes $T$ a \textit{planted}
tree\index{tree!planted} \cite[p.6]{TW-drm:09}.  An internal node
$\nu_j$ ($1\leq j < n$) marks the border between layers $l_j$ and
$l_{j+1}$ and layers subdivide any branch $b$ between two nodes into
branch \textit{segments}\index{branch!segment} $s(b)_1, \dots , s(b)_k$,
where $k$ depends on $b$.  The \textit{size}\index{branch!size} of a
branch is the number of leaves below the branch.  By extension, the
size of a segment\index{segment!size} is the size of the branch to
which the segment belongs.  A tree $T$ may be transformed into another
tree $\tilde T$ by a prune\index{subtree!pruning} and
re-graft\index{grafting!subtree}\index{subtree!grafting}
operation: (i) randomly select branch
segments $s_p$ in layer $l_p$ for pruning and $s_g$ in layer $l_g$ for
re-grafting, such that $l_g\leq l_p$; (ii) prune the subtree spanned
by $s_p$ and re-graft it to segment $s_g$.  This prune and re-graft
operation is a model of genetic
recombination\index{recombination}. Recombination can also be thought
of as a segmentation process, which subdivides a linear chromosome
into (genomic) segments\index{segment!genomic}, such that all sites
within one segment have the same genealogical history, or ranked tree; 
see the contributions of Baake and Baake~\cite{TW-EBMB20}, Birkner and 
Blath~\cite{TW-MBJB20} and Dutheil~\cite{TW-JD20} in this volume.
Here, we ask two questions: (i) what is the probability that
recombination changes the root of the tree and (ii) how is root
balance affected by recombination?  First note that only some
recombination events affect tree topology.  One way to change the root
is by a re-graft operation to a segment in layer $l_1$ above the root.
Such events may also change root balance $\omega_1$. Re-grafting below
the root may change root height or balance only if $s_p$ and $s_g$
belong to different root subtrees.  On average, this happens with
probability one third (see below).

So far, we ignored branch lengths, but for applications in population
genetics it is of interest to assign branch lengths according to the
coalescent process\index{process!coalescent}: the length of each layer
$l_j$ ($j>1$) is scaled by a factor proportional to $1/{j \choose 2}$.
Let $\tilde P_{\uparrow}(i)$ be the probability that a pruned branch
in such a coalescent tree\index{tree!coalescent} has size $i$ and that
re-grafting is above the current root, i.e., tree height
increases. Averaging over coalescent trees of size $n$, this
probability is \cite{TW-fdw:13}
\begin{equation}\label{TW-eq:p-i}
  {\tilde P}^{}_{\uparrow}(i)= \frac{2}{a^{}_n}\sum_{k=2}^n P^{}_{n,k}(i)
  \frac{1}{k(k-1)(k+1)}\,,
\end{equation}
where $a_n$ is the $n$-th harmonic number and
\begin{equation*}
P^{}_{n,k}(i)= \frac{{n-i-1 \choose k-2}}{{n-1 \choose k-1}} \,
\end{equation*}
is the probability that a branch of layer $k$ has size $i$.  Since
re-grafting is above the root, one of the root-subtrees will have size
$i$ after re-grafting and $\Omega_1$ will take the value
$\omega_1=\min(i,n-i)$ with probability
\begin{equation*}
P^{}_{\uparrow}(\omega^{}_1)=\frac{{\tilde P}^{}_{\uparrow}(\omega^{}_1)
+{\tilde P}^{}_{\uparrow}(n-\omega^{}_1)}{(1+\delta^{}_{2\omega_1,n})}\,.
\end{equation*}

Similarly, one can also obtain the transition probabilities from
$\omega_1^0$ before recombination to $\omega_1$ after recombination
when tree height is increasing.  Let $P_{n,j}(i\mid\omega_0)$ be the
probability that a branch at level $j$ has size $i$ in a tree of total
size $n$, given that the size of the root branches are $\omega_1^0$
and $n-\omega_1^0$. Then \cite{TW-fdw:13},
\begin{equation*}
  \tilde{P}^{}_{\uparrow}(i\mid\omega_1^0)= \frac{2}{a^{}_n}\sum_{j=2}^n P^{}_{n,j}
  (i\mid\omega_1^0) \frac{1}{j(j-1)(j+1)}
\end{equation*}
and
\begin{equation*}
P^{}_{\uparrow}(\omega^{}_1\mid\omega_1^0)  =  \frac{\tilde{P}^{}_{\uparrow}(\omega^{}_1\mid\omega_1^0)+\tilde{P}^{}_{\uparrow}(n-\omega^{}_1\mid \omega_1^0)}{1+\delta^{}_{2\omega^{}_1,n}}\,.
\end{equation*}
 
Similar calculations lead also to the transition probabilities of root
balance\index{balance!root} under recombination events that do not
change tree height, and to estimates of the `correlation length' of
root balance under multiple recombination events. These help to
explore the speed with which genealogical trees and shapes change
along a recombining chromosome.  Considering only recombination events
that change root height, we estimated the physical distance between
such recombination events as \cite[Eq.~(51)]{TW-fdw:13}
\begin{equation*}
\frac{1}{2(10-\pi^2)\rho}\sim \frac{3.83}{\rho}\,,
\end{equation*}
where $\rho$ is the scaled recombination rate per nucleotide site.  In
other words, about every $4$th recombination event affects tree
height. For example, if $\rho = 10^{-3}$, the genomic distance between
such events is about $4,000$ nucleotides.  Recombination events that 
affect root balance are slightly more common, since more branches are
available for re-grafting.  The distance between such events can be
estimated by the average run-length $(1-P(\omega_1\mid\omega_1))^{-1}$,
i.e., the average size of a genomic fragment within which root balance
$\omega_1$ does not change. The run-length depends on $n$, is longer
for more unbalanced trees (small $\omega_1$) and is on the order of a
few recombination events (about $2$ to $6$, for a typical sample size
of $n=100$) \cite{TW-fdw:13}.

\subsection*[Linkage disequilibrium]{Linkage disequilibrium}

Change in tree topology along a recombining chromosome can also be
interpreted as a reduction of \textit{linkage
  disequilibrium}\index{linkage disequilibrium}.  Two-loci linkage
disequilibrium, \textit{LD}, is the non-random association of two
alleles (genetic variants) from two linked genetic loci or sites
(alleles $A$, $a$ at the first locus and alleles $B$, $b$ at the
second, say). Let $X_{A}$ ($X_B$) be the indicator variable of allele
$A$ ($B$). Then, one standard way to express \textit{LD} is by
Pearson's correlation coefficient (e.g. \cite{TW-ZPW:08}) of the
indicator variables
\[
r^2 = \frac{\cov^2(X^{}_A,X^{}_B)}{\VV (X^{}_A) \VV (X^{}_B)}\,.
\]
Alleles $A$ and $B$ are often interpreted as being \textit{derived}
from their ancestral forms $a$ and $b$, respectively, by two
independent mutation events that occurred some time ago in the
genealogical history of each locus, i.e., by events that `fall on'
some branches of their genealogical trees.  As such, a mutation event
can be thought of as a `subtree marker', marking the subtree below the
branch on which it occurred. Thus, the frequency of the new mutation
in the current population(-sample) is identical to the size of the
marked subtree. Focusing on this property, one arrives at a slightly
modified concept of linkage disequilibrium \cite{TW-wrw:18}:
considering two, not necessarily adjacent, genomic segments, $S$ and
$U$, with \textit{labelled} ranked trees $T(S)$ and $T(U)$, and left
root subtrees\index{subtree!root} $T(S)_L$ and $T(U)_L$, the leaf
labels can be partitioned into four sets: (i) labels that belong to
both left subtrees, (ii) both right subtrees, (iii) to either the left
subtree in segment $S$ and right subtree in segment $U$, or (iv) vice
versa.  With the indicator variables $X_{T(S)_L}$ and $X_{T(U)_L}$ one
can calculate $r^2$ in exactly the same way as before and formulate
\begin{definition}\label{TW-def:tld}
The quantity 
\begin{equation*}
  r_{S,U}^2=\frac{\cov^2( X^{}_{T(S)^{}_L}, X^{}_{T(U)^{}_L} )}{\VV (X^{}_{T(S)^{}_L})
    \VV (X^{}_{T(U)^{}_L})}
\end{equation*}
is called \textit{topological linkage disequilibrium}\index{linkage disequilibrium!topological}
(\textit{tLD}) of the segments $S$ and $U$.
\end{definition}
Here, a segment takes the role of a gene locus, and left/right take
the roles of two alleles. The assignment of left and right is
arbitrary, as much as the naming of two alleles in the context of
conventional \textit{LD}, and does not affect the value
$r_{S,U}^2$. Let $S_L$, $S_R$, $U_L$ and $U_R$ denote the leaf labels
in the left and right root subtrees at segments $S$ and $U$. Note that
$r_{S,U}^2=1$, if and only if $S_L=U_L$ or $S_L=U_R$. This implies
that subtrees are not only identical in size but also contain 
identically labelled leaves at both segments. 

In contrast to conventional \textit{LD}, a configuration of complete
topological linkage, $r_{S,U}^2=1$, can be broken only by
recombination events that do change tree topology.  Since only about
every third recombination event changes tree topology, average decay
of \textit{tLD} with distance between segments is slower than decay of
conventional \textit{LD} \cite{TW-wrw:18}.  A simple argument is the
following: consider a pruning and a re-grafting event and the relative
size $p$ of the left root subtree.  The probability that both events
take place on opposite sides of the tree, i.e., on different root
subtrees, is $2p(1-p)$. Integrating with uniform density over all left
subtree sizes yields $\int_{p=0}^1 2 p (1-p) \dd p = 1/3$.  Furthermore,
\textit{tLD} has an about $3$-times higher signal-to-noise ratio (the
inverse of the coefficient of variation) than conventional \textit{LD}
\cite{TW-wrw:18}.  The limit of expected \textit{tLD} at large
distances between segments is
\[
\lim_{\rho\rightarrow  \infty}\EE \big( r_{S,U}^2(\rho)\big) = \frac{1}{n-1}\,,
\]
which is in agreement with a classical result by Haldane \cite{TW-hal:1940}. 

Generally, compared to conventional \textit{LD}, \textit{tLD} shows a
sharper contrast among genomic regions that are in low versus high
linkage disequilibrium. This is a welcome property when searching in
whole genome scans for signatures of potential gene-gene interactions
using patterns of linkage disequilibrium \cite{TW-wrw:18}.

\section[Evolving trees]{Transformations II: Pruning, grafting and
  evolving trees}\label{TW-section:evolving}
\subsection{The evolving Moran genealogy}

The Yule process\index{process!Yule} is a pure birth
process. Augmented by a death process, such that each split of a leaf
is compensated by removal of a uniformly chosen leaf and its parental
branch, size $n<\infty$ remains constant in time and the Yule process
becomes a \textit{Moran process}\index{process!Moran}.  Following the
Moran process over time $\tau$ naturally leads to the \textit{evolving
  Moran genealogy}\index{evolving!Moran genealogy}
$(\mbox{EMG}_{\tau})_{\tau\geq 0}$ (see the contribution of Kersting 
and Wakolbinger~\cite{TW-GKAW20} for a related class of evolving 
genealogies).  Conversely, for any time
$\tau=\tau^*$, a tree $T(\tau^*)$ of size $n$ can be extracted from the
sequence $(\mbox{EMG}_{\tau})_{\tau}$.  In the following, we consider
ordered, rather than un-ordered, trees and keep track of left/right
when choosing a leaf for splitting.

The evolving Moran genealogy, \textit{EMG} for short, induces a
discrete Markov process on the set $\LamO_n^{+-}$. This process is
recurrent and aperiodic \cite{TW-ww:19} and therefore has a stationary
distribution $P^*$ on $\LamO_n^{+-}$.  Since we may interpret the
genealogy $T(\tau)$ for any given $\tau$ as a result of a Yule
process, and since all $T$ are uniformly distributed, $P^*$ must be
the uniform distribution as well, i.e., $P^*(T)=1/(n-1)!$ (see
Table~\ref{TW-table:0}).

Following the process of tree balance\index{balance!tree} in an
\textit{EMG}, let $|T(\tau)_{L}|$ be the size of the left root subtree
of $T(\tau)$ extracted from $(\mbox{EMG}_{\tau})_{\tau}$.  The
sequence $(|T(\tau)_{L}|)_{\tau}$ is subject to the same transition
law as the frequency of a newly arising allele in a Moran model. A new
allele arising at time $\tau^*$ can be imagined as `marking' an
external branch of $T(\tau^*)$ and the evolving subtree under this
branch in $(\mbox{EMG}_{\tau})_{\tau > \tau^*}$. Only at the boundary,
there is an exception: whenever the left (or right) root subtree is of
size $1$, this remaining branch may be killed with positive
probability. This leads to loss or fixation of the allele and
consequently to a \textit{root jump}\index{root jump} with a uniform
`entrance' law. After a root jump the new left root subtree has
uniformly distributed size, and not necessarily size $1$.  We call the
time interval between successive root jumps an \textit{episode}\index{episode}
of the evolving Moran process.
\begin{result}
  For $2\leq |T(\tau)_L|\leq n-2$, the transition probability of the
  tree balance process $(|T(\tau)_{L}|)_{\tau}$ is given by
  \cite{TW-ww:19}
\begin{equation}\notag
\mathrm{Prob}\Big(|T(\tau+1)^{}_{L}|=\omega \mid |T(\tau)^{}_{L}|\Big)=
\begin{cases}
\frac{|T(\tau)^{}_{L}|(n-|T(\tau)^{}_{L}|)}{n^2}, & \omega = |T(\tau)^{}_{L}|+1\,, \\[2mm]
\frac{|T(\tau)^{}_{L}|^2+(n-|T(\tau)^{}_{L}|)^2}{n^2}, & \omega = |T(\tau)^{}_{L}|\,,\\[2mm]
\frac{|T(\tau)^{}_{L}|(n-|T(\tau)^{}_{L}|)}{n^2}, & \omega = |T(\tau)^{}_{L}|-1\,. \\
\end{cases}
\end{equation}
At the boundary $|T(\tau)_{L}|=1$, one has 
\begin{equation}\notag
\prob\left(|T(\tau+1)^{}_{L}|=\omega\right) =
\begin{cases}
\frac{1}{n}, & \omega = 2\,, \\[1mm]
\frac{(n-1)^2+2}{n^2}, & \omega = 1\,, \\[1mm]
\frac{1}{n^2}, & \textnormal{otherwise}\,, \\
\end{cases}
\end{equation} 
and at the boundary $|T(\tau)_{L}|=n-1$, one has
\begin{equation}\notag
\prob\left(|T(\tau+1)^{}_{L}|=\omega \right) =
\begin{cases}
\frac{1}{n}, & \omega = n-2\,, \\[1mm]
\frac{(n-1)^2+2}{n^2}, & \omega = n-1\,, \\[1mm]
\frac{1}{n^2}, & \textnormal{otherwise}\,. \\
\end{cases}
\end{equation} 
\end{result}
The result is proved by simple enumeration of the 
discretely many admissible events and calculation of their probabilities.

There is an alternative procedure of constructing an ordered ranked
tree of size $n$: by random grafting\index{grafting} of a new external
branch onto any branch segment of an existing tree of size
$n-1$. Random grafts\index{grafting} onto existing segments can take
place in two orientations, left- and
right-oriented\index{orientation}.  Both constructions are equivalent
and yield identical distributions. More precisely, we state the
following result.
\begin{result}
  The distributions of ordered Yule trees\index{tree!Yule} of size
  $n$, and those generated by successive random graftings are
  identical. Thus, for $T\in \LamO_n^{+-}$ generated by successive
  random graft operations from trees of size $n-1, n-2,\dots$, one has
$$
\prob (T)=\frac{1}{(n-1)!}
$$
\end{result}
The proof goes by induction on tree size and using a Lemma derived in
\cite{TW-ww:19}.

\subsection{Time reversal of the \textit{EMG}}

\begin{figure}
\begin{center}
 \includegraphics[scale=.35,trim=0 0 0 0,clip]{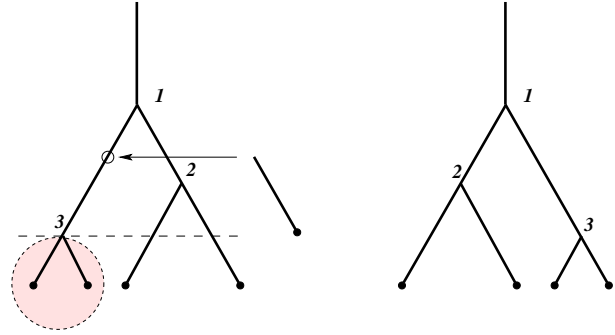}
\end{center}
\caption{Tree transformation by a merge-graft
  operation\protect\index{grafting} on the tree shown left. Removing
  the lowest layer (below the dashed line) leads to removal of the
  shaded cherry. A new branch is grafted (`resurrected') on a segment
  marked by the open circle. Labels are updated, resulting in the tree
  shown on the right. Both trees belong to $\LamO_4^{+-}$.}
\label{TW-figure:regraft}
\end{figure}

The Moran process\index{process!Moran} can be imagined as a
forward-in-time processes.  Reversing time, and starting with a
planted tree\index{tree!planted} $T\in \LamO_n^{+-}$ of size $n$,
consider now the following \textit{merge-graft} operation that 
generates a tree $T'\in \LamO_n^{+-}$: (i) including the branch
segment parental to the root, there are in total 
${n+1 \choose 2}$ segments in $T$; choose one branch segment $s^*$;
this is done with probability $1/n^2$ for segments ending in a leaf
($n$ possibilities) and with probability $2/n^2$ for all other
segments ($n\choose 2$ possibilities); (ii) if a leaf segment was
chosen then assign $T'\leftarrow T$. Otherwise, choose an
orientation\index{orientation} $\chi\in \{L, R\}$ (left/right) with
equal probability, remove the $n$-th layer from $T$, re-graft a new
branch in orientation $\chi$ at $s^*$, and update the labels of all
nodes.  The resulting tree is returned as $T'$ (see
Figure~\ref{TW-figure:regraft}).

Iterating the merge-graft operation\index{merge-graft operation} one
obtains a backward-in-time process that is dual to the Moran
process. We call the genealogy generated by this process the
\textit{evolving Moran genealogy backward in time}\index{evolving!Moran genealogy!backward in time}, 
$\textit{EMG}^\flat$ for short, and
note \cite{TW-ww:19} the following result.
\begin{result}
For all $T,T'\in \LamO_n^{+-}$:
\begin{equation}
  \prob^{}_{\textit{EMG}}\big(T(\tau+1)=T'\mid T(\tau)=T\big) =
  \prob^{}_{\textit{EMG}^\flat}\big({T}(\tau)=T\mid {T}(\tau+1)=T'\big)\,,
\label{TW-eq:transprob}
\end{equation}
with $\prob_{\textit{EMG}}$ ($\prob_{\textit{EMG}^\flat}$) denoting the
transition probability of the \textit{EMG}- (\textit{EMG}$^\flat$-)
process, respectively.
\end{result}

\subsection{The root jump process}
The $\textit{EMG}^\flat$ is interesting theoretically as well as
practically. While transitions in the $\textit{EMG}$ depend on two
random events, splitting and killing, in the $\textit{EMG}^\flat$
there is only one random operation, grafting\index{grafting}. This
fact simplifies some analytic approaches. Consider the root
jump\index{process!root jump} process.  An obvious question to ask is
how often do root jumps occur?  Working in the framework of the
$\textit{EMG}^\flat$, one can immediately state the following.
\begin{result}\label{TW-lemma:mrca1}
  Root jumps in the $\textit{EMG}$ and in the $\textit{EMG}^\flat$
  occur according to a geometric jump process of intensity
  $\frac{2}{n^2}$.
\end{result}
\begin{proof}
  In the $\textit{EMG}^\flat$, a root jump occurs if and only if the
  segment parental to the root is chosen for
  re-grafting. This happens with probability $\frac{2}{n^2}$. The same
  holds for the forward process due to duality.
\end{proof}
This result agrees with the one derived in \cite{TW-pw:06}, where the
jump process in the infinite-population limit is identified as a
Poisson process of intensity $1$. This is the limit of the geometric
jump process as $n\rightarrow \infty$ with time sped up by $n^2/2$,
which is the average number of Moran steps before a root jump occurs.
Also implied by Result~\ref{TW-lemma:mrca1}, the number of steps
needed to observe any number $k>0$ of root jumps follows a negative
binomial distribution with parameters $k$ and $2/n^2$.

Finally, using the simple structure of the $\textit{EMG}^\flat$, one
can calculate the number of root jumps during fixation of a new
allele. More precisely, consider time $\tau_0$ when a new allele $x^*$
is born (a subtree marker on some branch of $T$) and --- conditional on
fixation~--- time $\tau_1$ when $x^*$ becomes fixed, i.e., when all
leaves of $T(\tau_1)$ are descendants of $x^*$. We have the following
result.

\begin{result}\label{TW-lemma:fixjumps}
  In an \textit{EMG} of size $n\geq 2$, one expects
  $2 (1-\frac{1}{n})$ root jumps during the time interval
  $[\tau_0, \tau_1]$.
\end{result}
The proof goes by considering events in the backward process, where
one finds that the expected total number of root jumps along the
$\textit{EMG}^\flat$-path is
$$
\sum_{k=2}^{n-1}\frac{2}{k(k+1)}=\frac{n-2}{n}.
$$
Adding one additional jump, which necessarily happens at the moment of
fixation, one obtains the stated expectation.  Hence, in an infinitely
large sample ($n=\infty$) one expects two root jumps per one fixation,
a result obtained with different means before \cite{TW-pw:06}.

\begin{figure}
 \begin{center}
 \includegraphics[scale=1.3,trim=0 0 33 3,clip]{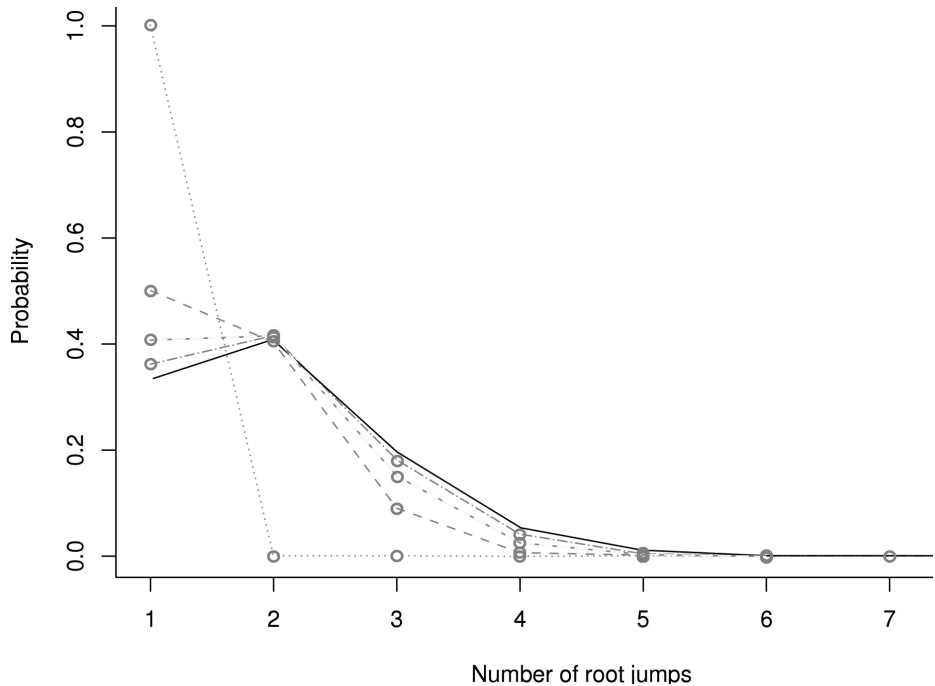}
 \end{center}
 \caption{The distributions of $P_n(k)$, $k=1,\dots,7$; $n=2$ (dotted), 
$n=5$ (dashed), $n=10$ (short dashes), $n=25$ (dot-dashed) and $n=\infty$ (black).}\label{TW-fig:rj}
\end{figure}

In the framework of the $\textit{EMG}^\flat$ one can calculate the
exact distribution of root jumps during a fixation recursively for any
$n$, and show that these distributions quickly converge as
$n\rightarrow \infty$. For $n\geq 2$, let $\prob_n(k)$ denote the
probability of observing $k$ root jumps during fixation of a new
allele in an $\textit{EMG}$ of size $n$, and $\prob_\infty(k)$ the same
probability in the infinite-population limit. Then,
$$
\prob^{}_n(k) = \sum_{2\leq i_1,\dots,i_{k-1}\leq n-1}
\prod_1^k\frac{2}{i^{}_k(i^{}_k+1)}
\prod_{j\neq i_1,\dots,i_{k-1}}\left(1-\frac{2}{j(j+1)}\right)\,.
$$
For small $k$, using software for symbolic algebra, one can easily
write down closed-form expressions for $\prob_n(k)$.  For $n=2$,
$\prob_2(1)=1$. $\prob_n(1)$ decreases monotonically in $n$, with
$\lim_{n\rightarrow \infty}\prob_n(1) = 1/3$.  In Figure~\ref{TW-fig:rj}
root jump distributions are shown for some small values of $n$ and for
$n=\infty$, illustrating the fast convergence for
$n\rightarrow \infty$.
 
Any root jump is tantamount to loss of some `genetic memory'. In the
future, it will be interesting to explore the root jump process in
more detail, in particular under non-equilibrium and non-neutral
population genetic scenarios, and with regard to the speed of loss of
genetic memory.

\subsection*{Acknowledgements}
I would like to express my gratitude to Filippo Disanto and Johannes Wirtz for
their intellectual input to the projects pursued as part of SPP~1590.
I am very grateful also to Luca Ferretti for his enthusiastic
discussions, sharing of ideas and his contributions to tree
transformations under recombination. Finally, I would like to thank
two reviewers for their critical and constructive comments on an
earlier version of this chapter.

\backmatter
\printindex

\end{document}